\documentclass[journal]{IEEEtran}
\usepackage[noadjust]{cite}
\usepackage{amssymb}
\usepackage{amsmath}
\usepackage{mathtools}
\usepackage{graphicx}
\usepackage{url}
\usepackage{siunitx}
\usepackage{amsmath} 
\usepackage{color}
\def\mod#1{\textcolor{black}{{#1}}} 

\DeclareMathOperator*{\argmin}{arg\,min}
\DeclareMathOperator{\Xop}{\mathcal{X}}
\DeclareMathOperator{\Dop}{\mathcal{D}}
\PassOptionsToPackage{obeyspaces}{url}
\graphicspath{{figs/}}
\mathtoolsset{showonlyrefs}

\title{Distributed optimization for nonrigid nano-tomography}
\author{Viktor~Nikitin,
Vincent~De~Andrade,
Azat~Slyamov,
Benjamin~J.~Gould,
Yuepeng~Zhang,
Vandana~Sampathkumar,
Narayanan~Kasthuri,
Do\u ga~G\" ursoy,~\IEEEmembership{Member,~IEEE},
Francesco~De~Carlo
\thanks{V. Nikitin, V. De Andrade, B. J. Gould, Y. Zhang, D. G\" ursoy, F.  De~Carlo are with Argonne National Laboratory, USA}
\thanks{A. Slyamov is with the Technical University of Denmark, Denmark}
\thanks{V.~Sampathkumar and N.~Kasthuri are with the Unviersity of Chicago, USA}
}

\begin{document}

\maketitle
\begin{abstract}
 Resolution level and reconstruction quality in nano-computed tomography (nano-CT) are in part limited by the stability of microscopes, because the magnitude of mechanical vibrations during scanning becomes comparable to the imaging resolution, and the ability of the samples to resist \mod{radiation induced deformations} during data acquisition. In such cases, there is no incentive in recovering the sample state at different time steps like in time-resolved reconstruction methods, but instead the goal is to retrieve a single reconstruction at the highest possible spatial resolution and without any imaging artifacts. Here we propose a distributed optimization solver for tomographic imaging of samples at the nanoscale. Our approach solves the tomography problem jointly with projection data alignment, nonrigid sample deformation correction, and regularization. Projection data consistency is regulated by dense optical flow estimated by Farneback's algorithm, leading to sharp sample reconstructions with less artifacts. Synthetic data tests show robustness of the method to Poisson and low-frequency background noise. We accelerated the solver on multi-GPU systems and validated the method on \mod{three} nano-imaging experimental data sets.
 \end{abstract}

\section{Introduction}
Current X-ray imaging instruments demonstrate sub-100\,nm resolution level in a wide range of fields, including biology, medicine, geology, and material sciences
~\cite{de2016nanoscale,Holler2017,holler2018omny,muller2018quantification,mizutani2019three,coburn2019design,gursoy2020multimodal}. To reach this or even higher resolution levels one needs to address various instrument component limitations such as mechanical instabilities from scanner, detector, and optic components. At the same time, the structure of a sample may change during X-ray data acquisition, leading to inconsistent tomographic data, further increasing the complexity of the sample reconstruction. This is caused not only by the controlled environmental conditions for \emph{in situ} study of dynamic processes, but also by the uncontrolled sample deformation due, for example, to radiation damage.
\mod{Severe radiation dose deposition may induce not only sample deformation but also ultimate destruction of features. Cryogenic systems like cryojet in most cases protect the sample from being destroyed, however, they} only partially reduce sample deformation while the cold air stream induces vibrations. On the other hand, in vacuum cryostages typically maintain samples at lower temperatures than cryostreams, more efficiently  preventing sample \mod{deformation} without adding vibrations. But such systems compatible with tomographic acquisitions are complex, unpractical and impose stringent experimental constraints ruling out most of \emph{in situ} experiments~\cite{wang2000soft,haibel2010latest,sorrentino2015mistral}. 

To address sample deformation due to radiation damage, a number of time-resolved methods have been adopted during the last two decades. They demonstrate significant quality improvement for time-evolving samples compared to the conventional approach, however, most of them are based on additional a priori knowledge about the sample structure and motion, and require demanding computational resources for reconstructing experimental data in a reasonable time. For instance, in~\cite{nikitin2019four} we proposed a multi-GPU implementation of a method for suppressing motion artifacts by using time-domain decomposition of functions with a basis chosen with respect to the motion structure. The method operates with a low number of decomposition coefficients that can be used to determine the object state continuously in time. 
Next, the approach proposed in~\cite{kazantsev20154d,kazantsev2016temporal} is based on estimating local structural correlations over multiple time frames and finding inner object edges which remain constant in time, followed by the patched-based regularization according to the object structure. Also, there are methods built upon the concept of compressed sensing, which employs sparsity promoting algorithms, where the prior knowledge is given in terms of spatial-temporal total variation regularization for detecting sharp object changes~\cite{wu2012spatial,ritschl2012iterative}. 

\mod{Compared} to the time-resolved methods above, the use of deformation vector fields (DVFs) is \mod{an even more} common approach to model the sample evolution at discrete time intervals during scanning. 
\mod{The displacement between volumes obtained from two independent sets of projections covering 180$^\circ$ intervals is generally recovered with the Digital Volume Correlation (DVC)~\cite{bay1999digital}. This works only if the deformation is slow compared to the scan speed, which is the case of some fast synchrotron imaging applications~\cite{rethore2011digital,keyes2017measurement,rossmann2020method}. Faster acquisition can be achieved by decreasing exposure times and the number of projection angles, however, DVC in this case suffers from noise and limited angle artifacts. 
In medical imaging applications, such as the study of heart or lungs, the displacement problem is easier to handle because the motion is often periodic and standard regrouping of projections yields reconstruction for different phases of the cycle~\cite{chen2011time,taubmann2015estimate}. However, the quality of reconstruction still suffers from the angular undersampling for each sample phase, so as the applicability is limited to a concrete deformation process.
Time-resolved DVC-based algorithms have also been proposed to deal with continuous sample deformation during data acquisition. The method proposed in \cite{neggers2015time} uses the sensitivity maps as the basis functions for representing DVFs, establishing a connection between the acquired images and the unknown material parameters. The DVC then operates only with these basis functions, allowing for faster deformation estimation. Direct material parameter identification provides strong regularization ensuring optimal accuracy and noise robustness. Another method \cite{leclerc2015projection} is based on the existence of the template sample volume recovered from high-quality undeformed projections. Further sample motion is recovered by using only a few projections per deformation. This approach can be efficiently used in dynamic imaging of samples with manual changing of temperature or pressure conditions, however, it will fail if some deformation happens during acquisition of the template. 
}

\mod{Several advanced methods have been recently proposed to address the above issues~\cite{odstrcil2019ab,de2018motion,zang2018space}. We see} two general limitations in using these methods for nano-CT applications with typically large-scale datasets. 
First, these methods are time-consuming even though they are implemented on GPUs. Because it is not feasible to store and process \mod{a large number of DVFs  for modeling rapid object changes}, the temporal resolution for the DVF estimation is chosen coarsely (e.g., one \mod{or several} time frames for each 180 degrees rotation)~\mod{\cite{ruhlandt2017four,odstrcil2019ab}}, \mod{so as the spatial resolution is decreased by applying binning (e.g., $4\times4$)} to get faster and more robust DVF estimation~\mod{\cite{odstrcil2019ab}}. \mod{Similarly, the authors in \cite{de2018motion} proposing the usage of B-splines for DVF estimation directly highlight that down-sampling can be carried out to perform GPU computations but at the cost of some loss in the displacement field’s accuracy. They also employ linear interpolation functions to estimate DVFs for time stamps corresponding to the acquisition of each projection, and claim that higher accuracy can be achieved by considering more complex functions. However, this will increase the computational complexity and memory requirements even more. An alternating joint optimization framework with computationally efficient DVF estimation was proposed in \cite{zang2018space}. With this approach the deformation field is estimated first for the coarsest scale, then the estimation is propagated step-by-step to the finest levels. According to the provided performance tests, a multi-threaded CPU implementation of the method requires 2-6 h to reconstruct 300$^3$-500$^3$ volumes with 6 alternating outer iterations, where most time is spent for DVF estimation even with the proposed efficient strategy. Taking into account the computational complexity and ignoring memory requirements these times may scale up to several days when considering volumes of twice bigger sizes.   }

\mod{The second general limitation is that} DVF estimation in 4D may result in many feasible solutions, and smoothness, slow-variations or other types of constraints are required to limit the solutions to obtain meaningful reconstructions. \mod{
Most algorithms are given without formal study of convergence analysis. Often, direct code usage for new experimental data results in divergence of the method  or in practically unacceptable results. In some cases, parameter adjustment and additional preprocessing procedures (binning, filtering, etc.) may improve reconstruction quality and make the method convergent. DVF-based methods typically operate with many parameters including initial and final DVF smoothness levels in spatial and temporal variables, regularizer for deformation evolution as in \cite{odstrcil2019ab}; weights $\kappa_1,\kappa_2,\kappa_3,\kappa_4$ for minimization terms as in \cite{zang2018space}. However, there is no general guidance on how to pick these parameters.}

In most experiments there is no requirement to reconstruct the object state at different time steps as proposed by computationally demanding time-resolved reconstruction methods, but instead the goal is to have a single reconstruction of high quality and resolution. These experiments include tomography of biological and medical samples, where X-ray dose deposition causes uncontrolled \mod{radiation or thermally} induced deformation~\cite{dyer2017quantifying,du2018relative,mizutani2019three,odstrcil2019ab}. Another application area is tomography of geological samples affected by external cooling or heating systems, where the sample structure changes depending on the rotation angle and corresponding distance to the external device~\cite{murshed2008natural,nikitin2020dynamic}. Even mechanical limitations of the rotation stage, such as inaccurate roll/pitch angles alignment or vibrations, in fact, create a time-resolved tomography problem where only one state needs to be recovered. 

The consistency of projections for different angles can be improved by directly translating projection images horizontally or vertically before the tomographic reconstruction~\cite{dierksen1993towards,castano2007fiducial,castano2010alignator,hayashida2010automatic,mizutani2019three}, or by using iterative re-projection methods based on a joint estimation of alignment errors and the object function~\cite{gursoy2017rapid,ramos2017automated,odstrvcil2019alignment}. These methods work especially well for the interlaced angular scanning protocol, in which a full tomographic scan is acquired with multiple sample rotations where projection angles are mod $2\pi$ different and uniformly cover the interval $[0,2\pi)$~\cite{mohan2015timbir, zang2018space}. With this protocol, a significant time gap between acquisition of two nearby angles, that should give similar data in a static case, allows to estimate corresponding shifts by using rigid image registration methods like cross-correlation.

In this work, we propose a generic distributed optimization framework based on alternating direction method of multipliers (ADMM)~\cite{Boyd:11} for solving the tomography problem jointly with \mod{two-dimensional} projected deformation estimation and regularization.
\mod{We operate with optical flow applied to projection data and not to the object, aiming to improve the  consistency  of  projections  and  recover  one  pristine  image of the sample  of high quality and resolution. This approach is computationally more appealing because 2D optical flow estimation is cheaper and can be straightforwardly parallelized over projections. The case with three-dimensional optical flow applied to the object is computationally demanding and well-suited for studying the dynamics of the sample, which is, however, not required for many nano-CT experiments.}
The proposed ADMM framework is favorable for handling large-scale problems because it breaks the whole reconstruction problem into algorithmically and computationally manageable sub-problems, in this case into three sub-problems: tomographic reconstruction, projected deformation estimation, and regularization of the solution. While the tomographic reconstruction and regularization sub-problems are solved using standard methods, we employ an optical flow scheme for estimating the projected deformation for each projection angle. Optical flow is defined as the pattern of apparent motion of image objects between two consecutive frames caused by the object movement, and is commonly used in computer vision for motion tracking algorithms~\cite{lucas1981iterative,farneback2003two,brox2010large}. 
\mod{We also provide a strategy for automatic parameter tuning in the ADMM scheme. So the algorithm can be run on different experimental data sets without additional adjustments except choosing the regularization constant. }
To our knowledge, a formal optimization framework has never been used for reconstruction problems with alignment and deformation corrections.  
In contrast to conventional re-projection methods relying on empirical studies of each sub-problem, the ADMM formulation connects all sub-problems and thus demonstrates more favorable convergence behavior, see~\cite{Boyd:11} and references therein. 
By estimating the optical flow in each projection image, one can handle a wide range of deformation types, even those occurring at fine time steps. 

Our implementation is based on Farneback's algorithm~\cite{farneback2003two} and we propose a new multi-resolution scheme for robust and accurate estimation of dense optical flow between projections. The algorithm can track the points at multiple levels of resolution by generating image pyramids, where each level has a lower resolution compared to the next (higher) level. Besides, in our proposed ADMM model, we solve all sub-problems approximately at each ADMM iteration and we use the estimates from the previous iteration as the initial guess for the next iteration. 
With these optimizations, processing times for the deformation estimation sub-problem \mod{are} comparable to that of the tomography sub-problem, and this strategy, combined with our multi-GPU code acceleration, allows processing large experimental data sets in a reasonable time.
We validated those optimizations both on synthetic and on large-scale experimental datasets. \mod{Reconstruction results of the proposed GPU implementation were compared to the ones obtained with a method operating with 3D optical flow~\cite{odstrcil2019ab}. Its publicly available Matlab GPU-based tomography solver with 3D optical flow estimation is available at \cite{odstrcil_michal_2019_2578796}.}

\mod{The proposed approach is a formalization and computational improvement of the method operating with the Projection-based Digital Volume Correlation (P-DVC) firstly introduced in~\cite{leclerc2015projection} where the theory was built upon the existence of the perfectly recovered initial object state.
The authors in~\cite{taillandier2016soft} demonstrate how P-DVC is used to recover sample changes based on the initial state with only a few radiographs acquired at fixed intervals of time. The ill-posedness of the inverse problem is addressed by incorporating an equilibrium-gap regularizer, registration is evaluated from the projection-based
residuals only.
In~\cite{jailin2018dynamic} tomographic reconstruction is directly coupled with the P-DVC problem in a way that two sub-problems are solved sequentially without coordinating variables. In addition, the method involves a multiscale Gaussian convolution procedure to properly correct displacements of different size and correct finer features. The technique was also used in online calibration procedures for cone-beam X-ray tomography~\cite{jailin2018self}. The authors claim that their method coupled with fast acquisition devices could give access to ultra-fast mechanical identification that could not be performed with classical means such as 3D or 4D DVC.
}

\mod{There is a straightforward applicability limitation of the proposed method and all methods based on optical flow estimation when reconstructing radiation sensitive samples. Radiation damage causes not only sample shrinkage and deformation but also destruction of features. With many features being destroyed by radiation, it becomes impossible to track optical flow between different states and, therefore, improve reconstruction quality and resolution. }

Our contribution through this work can be summarized as follows:
\begin{itemize}
    \item Development of a new ADMM-based optimization framework for joint tomographic reconstruction, geometrical data alignment, nonrigid sample deformation correction, and regularization;
    \item Use of a dense optical flow based scheme for jointly estimating the deformation fields of the projected images for each rotation angle to significantly suppress motion artifacts even for very small features;
    \item Algorithmic and computational optimizations for processing large-scale data in a reasonable time by use of multi-resolution, multi-GPU, and \mod{automatic} parameter tuning strategies;
    \item Demonstration of quality and spatial resolution enhancement when processing radiation-sensitive nano-CT experimental data.
    \item \mod{Comparison of reconstruction quality with an ad hoc GPU-based method operating with 3D optical flow.}
\end{itemize}

\mod{}

This paper is organized as follows. In Section~II, we formulate the joint tomography problem with deformation estimation and regularization, and show how this optimization problem can be solved with the ADMM by splitting the whole problem into sub-problems. In Section~III, we validate our approach on synthetic data and compare reconstruction results to the conventional method. In Section~IV, we provide performance optimization aspects when implementing the proposed method. Section~V demonstrates reconstruction results for nano-tomography experimental data sets. Our conclusions and outlook are given in Section~VI. 

\section{Methods}
In this section, we present the augmented Lagrangian formulation of the tomography reconstruction problem with deformation estimation and regularization, and show how this problem is solved by using ADMM~\cite{Boyd:11} with splitting the whole problem by local sub-problems. 

\subsection{Augmented Lagrangian formulation and solution by ADMM}
Let $u$ be an object in 3D space and $d$ its projection data around a common rotation axis. The projection operator is given in terms of the \mod{X-ray transform,
\begin{equation}
\mathcal{X}u(\theta,s,z)=\iint u(x,y,z)\delta(x\cos\theta+y\sin\theta - s)dxdy.    
\end{equation}}
For regularization, we use a common sparsity promoting total variation (TV) term  defined as 
\mod{
\begin{equation}
\alpha\|\nabla u\|_1=\alpha\left\lVert\sqrt{\big(\frac{\partial u}{\partial x}\big)^2+\big(\frac{\partial u}{\partial y}\big)^2+\big(\frac{\partial u}{\partial z}\big)^2}\right\rVert_1, \end{equation}}
where parameter $\alpha$ controls the trade-off between the data fidelity and regularization terms. We chose TV regularization as an example for suppressing noise and data incompleteness artifacts in reconstructions~\cite{Chambolle:16}, however other sparsity priors can be adopted as well.
Let $\Dop_f$ be the deformation operator that maps projections to new coordinates according to a given optical flow \mod{$f=(f_s^\theta,f_z^\theta)$ as follows, 
\begin{equation}
\Dop_f\psi(\theta,s,z) = \psi(\theta,s+f_s^\theta,z+f_z^\theta).
\end{equation}
Note that in this formulation we apply optical flow to projection data and not to the object. This way we aim at improving the consistency of projections to recover one pristine image of the sample with less motion artifacts. The case with three-dimensional optical flow applied to the object is computationally demanding and well-suited for studying the dynamics of the sample, which is not needed for experimental data we deal with in this work.}

The tomographic reconstruction problem of recovering object $u$ from data $d$ can be solved by considering the following minimization problem,

\begin{equation}
    \begin{aligned}
        \min_{u,f} \frac{1}{2}\|\Dop_f\Xop u - d\|^2_2+\alpha \|\nabla u\|_1.
    \end{aligned}
\end{equation}
\mod{We solve the above minimization problem by considering an equivalent constraint optimization problem where the augmented Lagrangian method is used to find a solution.}
By introducing auxiliary variables, $\psi_1$ and $\psi_2$, an equivalent problem can be written as a constrained optimization problem,
\begin{equation}\label{Eq:minF}
\begin{aligned}
    & \min_{u, f, \psi_1, \psi_2}     & & \frac{1}{2}\|\Dop_f \psi_1-d\|_2^2+\alpha \|\psi_2\|_1  \\
    & \text{subject to}        & & \Xop u = \psi_1, \\
    &                          & & \nabla u = \psi_2.
\end{aligned}
\end{equation}
The augmented Lagrangian for the problem \eqref{Eq:minF} is defined for real variables and is given as
\begin{equation}\label{Eq:alagr}
\begin{aligned}
  &\mathcal{L}_{\rho_1, \rho_2} (u,f,\!\psi_1,\psi_2, \lambda_1,\lambda_2)=\\&\frac{1}{2}\|\Dop_f \psi_1-d\|_2^2 + \lambda_1^T(\Xop u-\psi_1)+\alpha \|\psi_2\|_1+\\ & \frac{\rho_1}{2}\|\Xop u-\psi_1\|_2^2+ \lambda_2^T(\nabla u-\psi_2) + \frac{\rho_2}{2}\|\nabla u-\psi_2\|_2^2,
\end{aligned}
\end{equation}
where $\rho_1,\rho_2>0$ are penalty parameters and $\lambda_1,\lambda_2$ denote dual variables. 
We use ADMM to split the minimization problem of the augmented Lagrangian into three local sub-problems with respect to $u,(f,\psi_1),\psi_2$. The sub-problems are then coordinated through variables $\lambda_1,\lambda_2$ to find a solution for the original problem. Specifically, the following steps are performed in each ADMM iteration $k$:
\begin{align}
  &u^{k+1} =      \argmin_{u}      \mathcal{L}_{\rho_1,\rho_2}\!\left(u,\psi_1^{k},\psi_2^{k},\lambda_1^{k},\lambda_2^{k},f^{k} \right)\!,\label{Eq:tomo}\\
  &\psi_1^{k+1}\!,f^{k+1} =  \argmin_{\psi_1,f} \mathcal{L}_{\rho_1,\rho_2}\!\left(u^{k+1},\psi_1,\psi_2^k, \lambda_1^{k},\lambda_2^{k},f^k \right)\!,\label{Eq:deform}\\
  &\psi_2^{k+1} =  \argmin_{\psi_2} \mathcal{L}_{\rho_1,\rho_2} \left(u^{k+1},\psi_1^{k+1},\psi_2, \lambda_1^{k},\lambda_2^{k},f^{k+1} \right)\!,\label{Eq:regul}\\
  &\lambda_1^{k+1} =  \lambda_1^k + \rho_1 \left(\Xop u^{k+1} - \psi_1^{k+1}\right)\!,\label{Eq:lamd1}\\
  &\lambda_2^{k+1} = \lambda_2^k + \rho_2 \left(\nabla u^{k+1} -\psi_2^{k+1}\right)\!,\label{Eq:lamd2}
\end{align}
for zeros or some adequate initial guess at $k=0$. 
On every iteration, $\mathcal{L}_{\rho_1, \rho_2}$ is minimized over each variable sequentially by using the most recent updates of other variables. The dual variables are then computed with respect to the scaled sum of the consensus errors. \mod{With dropping terms not affecting minimization, the sub-problems \eqref{Eq:tomo}, \eqref{Eq:deform}, and \eqref{Eq:regul} can be written as 
\begin{align}
 &u^{k+1} = \lambda_1^T(\Xop u-\psi_1^{k})+ \frac{\rho_1}{2}\|\Xop u-\psi_1^k\|_2^2\\&\hspace{2.85cm}+ \lambda_2^T(\nabla u-\psi_2^k) + \frac{\rho_2}{2}\|\nabla u-\psi_2^k\|_2^2,\label{Eq:tomo2}\\
 &\psi_1^{k+1}\!,f^{k+1} = \frac{1}{2}\|\Dop_f \psi_1-d\|_2^2 +\lambda_1^T(\Xop u^{k+1}-\psi_1)\\&\hspace{4.7cm}+\frac{\rho_1}{2}\|\Xop u^{k+1}-\psi_1\|_2^2,\label{Eq:deform2}\\
 &\psi_2^{k+1} = \alpha \|\psi_2\|_1+ \lambda_2^T(\nabla u^{k+1}-\psi_2)+ \frac{\rho_2}{2}\|\nabla u^{k+1}-\psi_2\|_2^2.\label{Eq:regul2}
\end{align}}
In what follows, we show our solution approach to the sub-problems in~\eqref{Eq:tomo2},~\eqref{Eq:deform2}, and~\eqref{Eq:regul2}. We refer to them as the tomography, deformation estimation, and regularization sub-problems, respectively.

\subsection{Tomography sub-problem}

We express the minimization function for the tomography problem in \eqref{Eq:tomo2} as
\begin{equation}\label{Eq:tomoe}
  F(u)= 
  \frac{\rho_1}{2}\|\Xop u-\psi_1^{k}+\lambda_1^k/\rho_1\|^2_2+ \frac{\rho_2}{2}\|\nabla u-\psi_2^k+\lambda_2^{k}/\rho_2\|_2^2,
\end{equation}
where the terms not depending on $u$ are dropped. 
The problem is solved by considering the steepest ascent direction $\nabla_u F(u)$ computed as follows,
\begin{equation}
  \nabla_u F(u) = \rho_1\!\Xop^T\!(\Xop u -\psi_1^k+\lambda_1^k/\rho_1)-\rho_2\text{div}(\nabla u -\psi_2^k+\lambda_2^k/\rho_2),
  \label{Eq:gradtomo}
\end{equation}
where the divergence operator $\text{div}$ is the adjoint to $-\nabla$\mod{ and the adjoint operator for the X-ray transform, $\mathcal{X}^T$, is defined as 
\begin{equation}
\mathcal{X}^T\psi(x,y,z)=\iint \psi(\theta,s,z)\delta(x\cos\theta+y\sin\theta - s)d\theta ds.
\end{equation}}
With the steepest ascent direction, we can construct iterative schemes for solving \eqref{Eq:tomo2} by using methods with different convergence rates. Here we employ the Conjugate Gradient (CG) method for its faster convergence rate at the expense of memory requirements. CG iterations are given as $u_{m+1} = u_m + \gamma_m \eta_m$,
where $\gamma_m$ is a step length computed by a line-search procedure~\cite{nocedal2006numerical} and $\eta_m$ is the search direction that we compute by using the Dai-Yuan formula~\cite{DaiYuan:99},
\mod{
\begin{equation}\label{Eq:DaiYuan2}
\begin{aligned}
  \eta_{m+1}\!=\!-\nabla_u F(u_{m+1})\!+\!\frac{\|\nabla_u F(u_{m+1})\|_2^2}{(\nabla_u F(u_{m+1})\!-\!\nabla_u F(u_{m}))^T\eta_m}\eta_m,
  \end{aligned}
\end{equation}
where $\eta_0=-\nabla_u F(u_0)$.}
On each ADMM iteration, we solve the tomography sub-problem approximately, by using only a few number of CG iterations since \mod{this strategy in practice} greatly improves ADMM convergence rates, see review in~\cite{Boyd:11} and Section~IV. \mod{Note that solving the tomography sub-problem is always coordinated with solving other sub-problems}.

\subsection{Deformation estimation sub-problem}\label{Sec:def}
The minimization functional for the deformation estimation sub-problem in \eqref{Eq:deform2} can be expressed as
\begin{equation}\label{Eq:minFdef}
    F(\psi_1,f) = \frac{1}{2}\|\Dop_f \psi_1-d\|_2^2+\frac{\rho_1}{2}\|\Xop u^{k+1}-\psi_1+\lambda_1^k/\rho_1\|_2^2.
\end{equation}
As for the tomography sub-problem, we aim at solving this problem approximately on each ADMM iteration. Therefore, we decided not to increase the problem complexity by introducing additional coordinating variables with respect to $\psi_1$ and $f$. Instead, we propose sequential updates of these variables while solving the deformation estimation sub-problem.  
To find the optical flow we fix variable $\psi_1$ and solve
\begin{equation}
    \tilde{f} = \argmin_f \frac{1}{2}\|\Dop_f \psi_1-d\|_2^2.
\end{equation}
Sparse and dense optical flow estimation methods have been proposed by many authors, including Lucas and Kanade~\cite{lucas1981iterative}, Farneback~\cite{farneback2003two}, Brox~\cite{brox2010large}.  
In this work we adopted Farneback's algorithm because it gives accurate dense flow estimation, where the density level is controlled through parameters. Moreover, the algorithm allows solving the problem approximately (for a few iterations), which is favorable for the proposed ADMM scheme and consistent with the tomography sub-problem. At the same time, the choice of the method generally depends on the application, and other optical flow or other types of nonrigid registration algorithms can be applied as well.

Farneback's algorithm generates an image pyramid, where each level has a lower resolution compared to the next (higher) level. The scheme can track the points at multiple levels of resolution, starting at the lowest level typically given by a window size parameter that is close to the image size. Increasing the number of pyramid levels enables the algorithm to handle larger displacements of points between frames. However, the number of computations also increases with increasing resolution levels.

In the proposed reconstruction scheme there is no need to consider many pyramids levels for tracking deformation, because a level of three to five can capture most of the deformation sizes and types that are visible in real applications. Variable $\psi_1$ is updated on each ADMM iteration with the usage of the current object approximation $u^{k+1}$ and dual variable $\lambda_1^k$ that are typically not very close to final results on the first ADMM iterations. Thus at initial iterations it makes sense to find optical flow in low resolution with a large window size, and gradually increase resolution by decreasing the window size with iterations. Moreover, since Farneback's algorithm is an iterative procedure, the optical flow on each ADMM iteration can be used as an initial guess for the flow in the next iteration to accelerate solution. Therefore, the number of iterations in the algorithm can be kept low.

The problem with respect to variable $\psi_1$ with fixed $f$,
\begin{equation}
   \tilde{\psi}_1=\argmin_{\psi_1}\frac{1}{2}\|\Dop_f \psi_1-d\|_2^2+\frac{\rho_1}{2}\|\Xop u^{k+1}-\psi_1+\lambda_1^k/\rho_1\|_2^2,
\end{equation}
is treated as a standard $L_2$ minimization problem with linear operators, \mod{where the steepest ascent direction is given as follows,
\begin{equation}
  \nabla_{\psi_1} F(\psi_1) = \Dop_f^T(\Dop_f \psi_1-d)+\rho_1\Xop^T(\Xop u^{k+1}-\psi_1+\lambda_1^k/\rho_1).
\end{equation}}
For a consistent approach with the tomography problem, we employ the CG method with the Dai-Yuan search direction \eqref{Eq:DaiYuan2} to find a solution.

\subsection{Regularization sub-problem}

The minimization functional for the regularization problem \eqref{Eq:regul2} is given by
\begin{equation}\label{Eq:regproblem}
  F(\psi_2) = \alpha \|\psi_2\|_1 + \frac{\rho_2}{2}\|\nabla u^{k+1}-\psi_2  + \lambda_2^k/\rho_2\|_2^2.
\end{equation}
\mod{As opposed to the tomography and deformation solvers involving iterative solution approximation, the minimizer for $F(\psi_2)$ can be found} explicitly by using the soft-thresholding operator~\cite{Donoho:95},
\begin{equation}
  \tilde{\psi}_2 = \frac{\nabla u^{k+1}+2\lambda_2^k/\rho_2}{|\nabla u^{k+1}+2\lambda_2^k/\rho_2|}\max(0,|\nabla u^{k+1}+2\lambda_2^k/\rho_2|-2\alpha/\rho_2),
  \label{Eq:solreg}
\end{equation}
where, by abuse of notation, we assume element-wise operations for dividing, taking the absolute value, and comparing with 0.

We have considered a sparse prior defined in terms of the total variation, which is a common method of choice for reconstructing data from noisy and incomplete measurements. 
Although we use TV for regularization, the proposed framework allows the use of any other regularization approach based on the experimental conditions. For instance, one can use $\|\nabla u\|_2$ for decreasing the gradient sparsity level, \mod{or consider TV modifications such as the Huber variant of
the total variation prior (HTV)~\cite{huber2004robust}, the total generalized variation (TGV)~\cite{bredies2010total}. Alternatively, regularization schemes based on a sparse data representation~\cite{lee2001wavelet,colonna2010radon,cerejeiras2011inversion} can be adopted for the proposed scheme as well.}

\mod{Different techniques exist for choosing the regularization constant $\alpha$ in~\eqref{Eq:regproblem}. The most popular of them include the discrepancy principle~\cite{Wen:11}, L-curve criterion~\cite{Agarwal:03}, and generalized cross-validation~\cite{Golub:79}. In this work we use the L-curve criterion since it demonstrates robustness to noise and sharp solutions in our applications.
The L-curve criterion seeks to balance the error components $\frac{1}{2}\|\Dop_f\Xop u - d\|^2_2$ and $\alpha\|\nabla u\|_1$ via inspection of the L-curve constructed by using logarithms of these two errors. It has two distinctly different parts: quite flat (regularization error dominates), and more vertical (fidelity error dominates). The region where one part changes to another then corresponds to optimal values of the regularization parameter $\alpha$.  
}

\mod{
\subsection{Convergence note}
The constrained optimization problem \eqref{Eq:minF} is non-convex. Unlike the convex case, for which the behavior of ADMM has been studied extensively, the convergence of ADMM on non-convex problems is always questionable, especially when there are also non-smooth functions in the problems. The  ADMM  scheme  generally  fails  on  non-convex  problems, however it  has  not only demonstrated convergent results in many applications but also showed significant performance improvement. Indeed, a list of successful applications includea phase retrieval \cite{wen2012alternating}, 3D ptychography~\cite{Aslan:19,nikitin2019photon}, compressed sensing~\cite{chartrand2013nonconvex}, background/foreground extraction~\cite{yang2017alternating}, image registration~\cite{bouaziz2013sparse}.
Most of these applications are far beyond the scope of theoretical conditions required for non-convex ADMM convergence~\cite{wang2019global,hong2016convergence,wang2018convergence,xu2012alternating}. Nevertheless, they are still practically used in real data processing.}

\mod{
For a detailed convergence discussion of the ADMM scheme for the minimization problem studied in this paper we refer readers to~\cite{hong2016convergence}. The authors analyze the following problem,
\begin{equation}    \label{Eq:hongADMM}
    \begin{aligned}
    & \min_{x_1,x_2,\dots x_K} \sum_{k=1}^K g_k(x_k) + l(x_1,x_2,\dots, x_K)\\
    & \text{\qquad s.t.\,\, } \sum_{k=1}^K A_k x_k = 0,
    \end{aligned}
\end{equation} 
where $A_k$ are linear operators, $l(\cdot)$ is a smooth
(possibly non-convex) function,  and each $g_k$ can be either a smooth function or
a convex non-smooth function.  
The problem~\eqref{Eq:minF} has a form of~\eqref{Eq:hongADMM} if we set variables and functionals as
\begin{align}
    &x_1=\psi_1,\quad x_2=\psi_2,\quad x_3=u,\quad x_4=f,\\
    &g_2(x_2)=\alpha\|x_2\|,\quad l(x_2,x_4) = \frac{1}{2}\|\Dop_{x_4}x_2-d\|_2^2,
\end{align}
and the operators in the constraint as
\begin{align}
    A_1 = \begin{pmatrix}I\\\mathbf{0}\end{pmatrix}, \, A_2 = \begin{pmatrix}\mathbf{0}\\I\end{pmatrix}, A_3 = \begin{pmatrix}-\Xop\\-\nabla\end{pmatrix}.
\end{align}}

\mod{
The rate of convergence depends on the choice of the ADMM weighting parameters $\rho_1,\rho_2$. When these parameters are well chosen, the method can converge to fairly accurate solutions within a few tens of iterations. However, if the choice of the parameters is poor, many iterations are typically needed for convergence. Note that the case with $\rho_1,\rho_2=0$ corresponds to a standard  alternating approach for solving the tomography problem with deformation estimation (used, for instance, in~\cite{odstrcil2019ab}) where two sub-problems are solved sequentially without coordinating $L^2$-norm terms weighted with  $\rho_1,\rho_2$ as in \eqref{Eq:tomo2}-\eqref{Eq:regul2}, and therefore demonstrates less favorable convergence behavior in practice. Heuristics for choosing optimal penalty parameters are discussed more in detail in \cite{Boyd:11}, and references therein. In Section IV we also describe the strategy of automatic updating these parameters for a faster convergence.
}
\section{Synthetic data tests}
\begin{figure}[htb]
    \centering
    \includegraphics[width=0.48\textwidth]{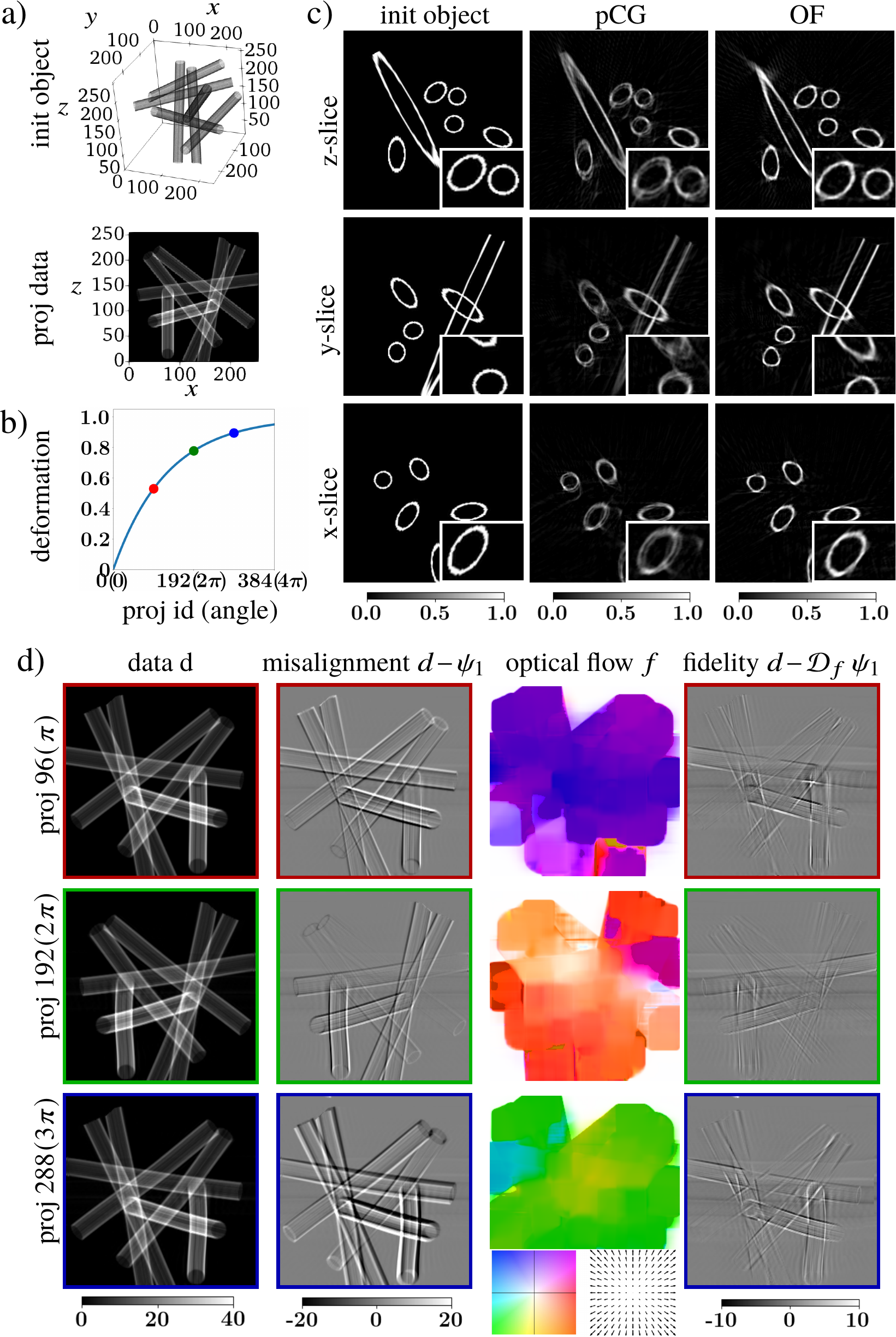}
    \caption{Reconstruction of a deforming synthetic object by the Conjugate-Gradient method with pre-alignment before reconstruction {(pCG)}, and by the proposed ADMM scheme with optical flow {(OF)}: a) initial object with an example of projection data, b) deformation speed plot, c) orthogonal slices through the initial object and reconstructions, d) application of the deformation operator $\Dop_f$ with the recovered optical flow $f$ to the re-projection data $\psi_1$: misalignment and fidelity terms with respect to the data $d$ acquired at the object states marked by colored dots in the plot from b).}
    \label{Fig:figinit}
\end{figure}

In this section, we validate our approach through simulations. We will compare reconstructions by the standard Conjugate Gradient solver with pre-aligned projection, referred to as \textbf{pCG}, and by the proposed ADMM scheme with optical flow, referred to as \textbf{OF}. Noise and additional background robustness for both methods is analyzed with adding the TV regularization term. TV-regularized versions for both methods, pCG and OF, are referred to as \textbf{pCGTV} and \textbf{OFTV}, respectively. 

For validation we generated projection data for a continuously deforming synthetic object represented by a set of tube-like structures of different sizes and orientation, see Fig.~\ref{Fig:figinit}(a). The amount of deformation for a normalized scanning time $t\in[0,1]$ is proportional to $1-
\text{exp}(-3t)$, see Fig.~\ref{Fig:figinit}(b). The plot approximates a slowing down process of the object deformation caused by radiation damage during X-ray scanning. The deformation was simulated as a smooth random field with maximal displacement of 20 pixels (8\% of the object size in one dimension).
In total 384 projections through the $256\times256\times256$ object were acquired in the interlaced scanning mode with 2 object rotations. 

Fig.~\ref{Fig:figinit}(c) shows reconstruction results comparison for the pCG and OF methods. In all presented orthogonal slices through the reconstructed object, the OF method suppress deformation artifacts appeared in pCG reconstructions as blurred and corrupted ellipses. Note that the OF method recovers not the initial object state, but the state which yields a higher consistency level of the projection data.       

Projections in the pCG method were pre-aligned with respect to the projections from the first rotation. Instead of the conventional cross-correlation approach for alignment, we employed Farneback's algorithm with non-dense optical flow. This approach also gives one common $(x,y)$-shift for all pixels in a projection, but it is more robust to noise than the cross-correlation approach. Alternatively, one could consider dense optical flow for pre-alignment, however, this results in even worse reconstructions because interlaced projections were acquired for different angles and thus are not supposed to be the same up to small features.

For the OF method, in turn, projections were aligned with respect to the re-projection estimate $\psi_1$, see \eqref{Eq:minF}. In Fig.~\ref{Fig:figinit}(d) we show how recovered estimates $\psi_1$ differ from the data $d$ acquired at different time steps marked by colored dots in Fig.~\ref{Fig:figinit}(b). We also show that application of the estimated optical flow $f$ to the re-projection $\psi_1$ minimizes the data fidelity error $d-\Dop_f\psi_1$. Visualization of the optical flow is done by using a colored map where colors define flow directions, and color intensities show the amount of deformation.

The iterative scheme for the pCG method was performed for 128 iterations, whereas the OF scheme had 64 outer ADMM iterations and 4 inner iterations for each tomography and deformation estimation sub-problems (\mod{using of few inner iterations} is a common practice to speed up ADMM). The optical flow density level was gradually increased with iterations by reducing window sizes in Farneback's algorithm from 256 to 32 pixels.

\begin{figure}[htb]
    \centering
    \includegraphics[width=0.48\textwidth]{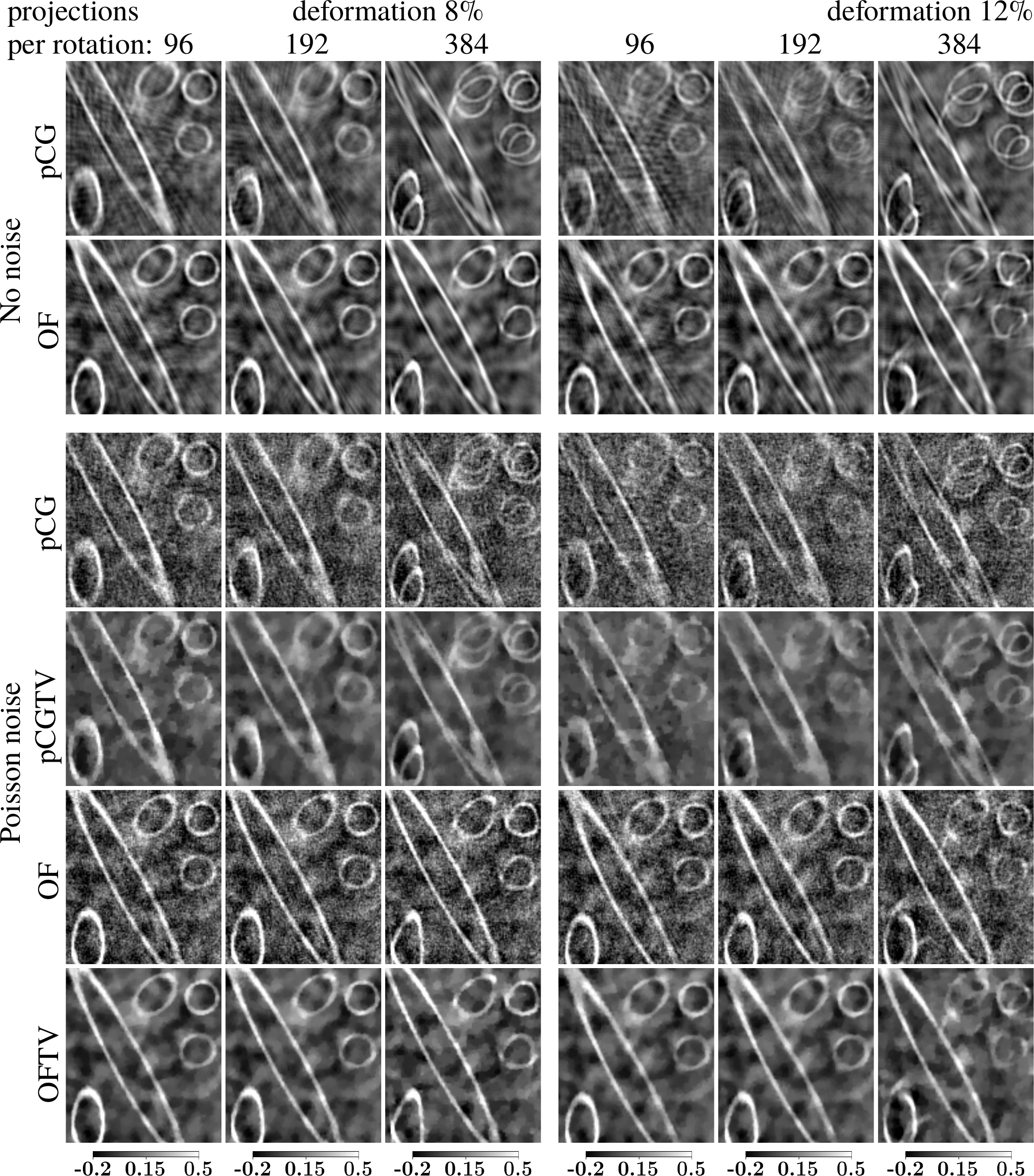}
    \caption{Reconstruction of a deforming synthetic object from the data with Poisson and low-frequency background noise by using the proposed method OF, by the conventional pCG, and by their regularized versions, OFTV and pCGTV. }
    \label{Fig:fignoise}
\end{figure}

To check the noise robustness and stability of the proposed method to background changes we modified the synthetic object from Fig.~\ref{Fig:figinit}(a) by varying intensities of the tube-like structures, as well as adding Poisson noise and low-frequency background noise to projection data. We also considered two levels of deformation simulated as smooth random fields with maximal displacement of 20 and 30 pixels (8\% and 12\% of the object size), and analyzed how the number of angular rotations in the interlaced scan affects reconstruction quality. Results are given in Fig.~\ref{Fig:fignoise}, where for clarity we present cropped z-slices through the reconstructed object. There are several observations from this figure.
The OF method is more robust to Poisson noise and low-frequency background noise than the pCG method. Additional TV regularization reduces high-frequency noise in OFTV reconstructions, whereas pCGTV reconstructions \mod{blur main object features}. Next, the lower number of projections per rotation in the interlaced scanning mode is favorable for the proposed method but not always for the conventional method. One can observe straight line artifacts in pCG reconstructions for the case with 96 projections per rotation. The artifacts appeared because the alignment was done between interlaced projections that are rather more different than in the case with 192 and 384 projections. We also note that the proposed method partially resolved deformation issues for the non-interlaced scanning (1 rotation with 384 angles).  

In practice, lowering the number of projections per rotation introduces additional overhead for rotating the stage, which leads to longer acquisition times and skipping some of the deforming sample states. Although the proposed method significantly suppresses motion artifacts, the trade-off between the rotation speed and reconstruction quality depends on the sample dynamics and needs to be determined experimentally.

\section{Performance optimization}\label{Sec:optim}

In this section we describe optimization aspects for implementing the proposed ADMM scheme in order to use it in large-scale experimental data processing.
\begin{figure*}[htb!]
    \centering
    \includegraphics[width=0.75\textwidth]{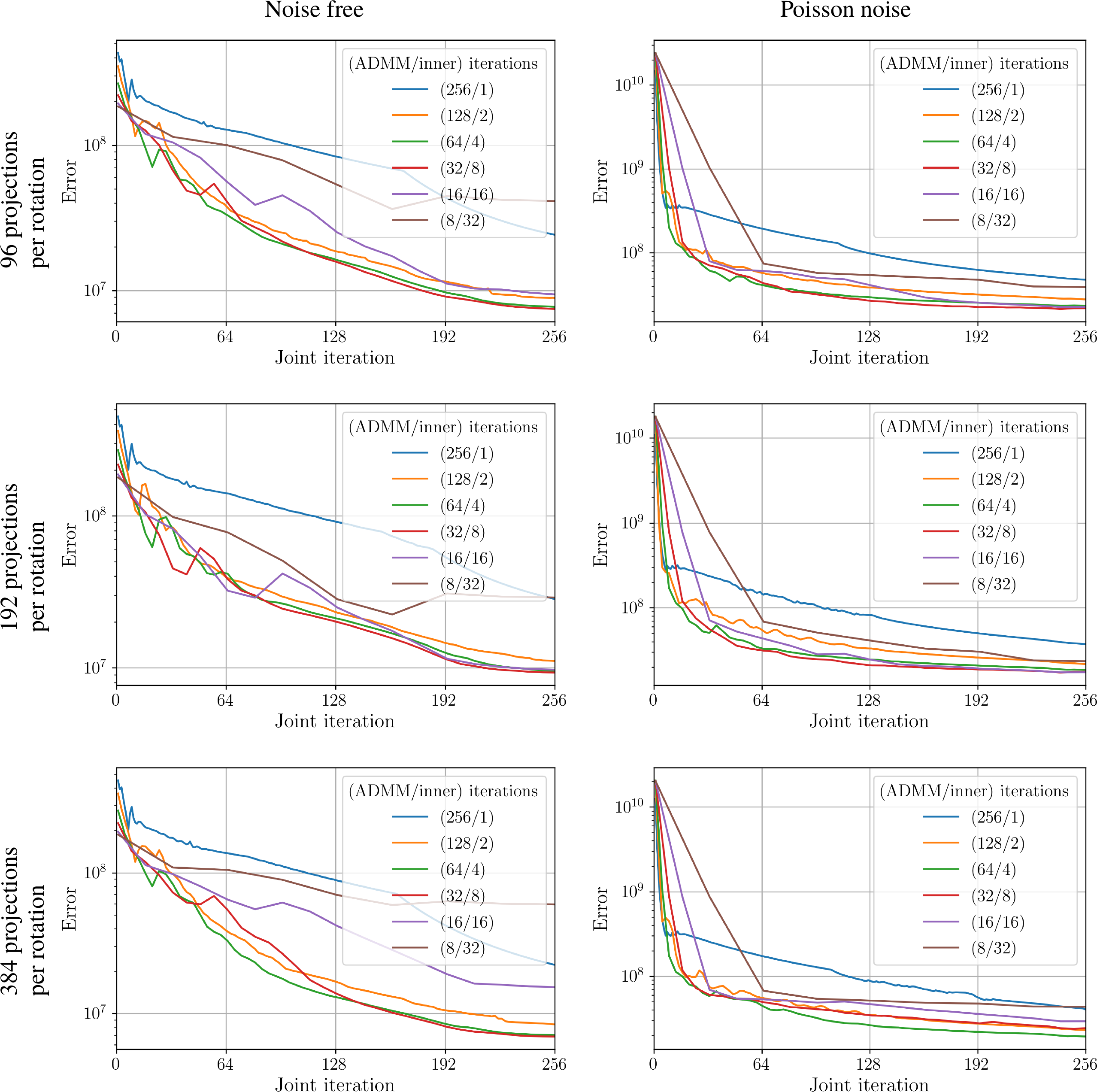}
    \caption{Convergence plots for the augmented Lagrangian function minimized by the proposed scheme with $n_{\text{inner}}$=1,2,4,8,16,32 inner CG iterations in the tomography and deformation estimation sub-problems. Number of global ADMM iterations $n_\text{ADMM}=256/n_\text{inner}$ yields the same computational complexity level for all cases. Joint iteration is calculated as the product of the ADMM iteration count and $n_{\text{inner}}$.  }
    \label{fig:convergence}
\end{figure*}
\begin{table*}[htb!]
\caption{Computational times (in seconds) of 1 ADMM iteration with 4 inner CG iterations for tomography and deformation estimation sub-problems. The GPU system for testing is based on 4$\times$NVIDIA Tesla P100. }
    \centering
    \def\arraystretch{1.5}
    \begin{tabular}{|c|c|c|c|c|c|c|}
    \hline
        Data size & \multicolumn{2}{c|}{Deformation estimation} & \multicolumn{2}{c|}{Tomography} & \multicolumn{2}{c|}{Total}  \\\cline{2-7}
        ($N_\theta\!\times\!N_z\!\times\!N$)& 1 GPU & 4 GPU & 1 GPU & 4 GPU & 1 GPU & 4 GPU \\ \hline
         $196\!\times\!128\!\times\!128$ & 0.9 & 0.7 & 0.5 & 0.6 & 1.4 & 1.3    \\
         $384\!\times\!256\!\times\!256$ & 2.2 & 1.4 & 1.6 & 1.4 & 3.9 & 3.0  \\
         $768\!\times\!512\!\times\!512$ & 12.7 & 7.4& 18.3 & 8.4 & 32.3 & 17.2  \\
         $1536\!\times\!1024\!\times\!1024$ & 105 & 63 & 197 & 107 & 315 & 183\\
         $3072\!\times\!2048\!\times\!2048$ & 897 & 598 &1976&1131&3001 &1810          \\
         \hline
    \end{tabular}
    \label{tab:perf}
\end{table*}

Convergence behavior and performance of the scheme were optimized with the following aspects.
First, on each ADMM iteration we propose solving the tomography and deformation estimation sub-problems approximately, by using only a few number of inner iterations.
In practice, this procedure yields more favorable convergence behavior than the one with many inner iterations and allows to significantly speed up the whole ADMM scheme~\cite{Aslan:19,nikitin2019photon}. In Fig.~\ref{fig:convergence} we present convergence results for different number of inner CG iterations in the tomography and deformation estimation sub-problems. The figure demonstrates convergence plots of the augmented Lagrangian function for the schemes having the same computational complexity level in the sense that the product of the total number of ADMM iterations and the number of inner CG iterations is the same for all schemes. The tests were performed for synthetic 10\% deformed data used in the paper. According to this figure, solving the problem with a low number of inner iterations (specifically with 4) demonstrates more favorable convergence results compared to other cases.

Second, we work with normalized and computationally optimized operators defined in the paper for constructing the ADMM framework. This facilitates the initialization of ADMM weighting factors $\rho_1$ and $\rho_2$ in the augmented Lagrangian and optimize line-search computations in the CG method for the tomography and deformation estimation sub-problems. Additionally, we follow~\cite{Boyd:11} and vary the penalties $\rho_1,\rho_2$ on each iteration to improve the convergence rate. 
\mod{
With this strategy, the update of $\rho_1$ for each iteration $k$ is computed as follows
\begin{equation}\label{Eq:penalty_update}
\begin{aligned}
  &\rho_1^{k+1}=\\&\begin{cases}
    2\rho_1^k    & \text{if } \|\psi_1^k-\Xop u^k\|^2_2 > 10 \|\rho_1^k(\Xop u ^{k}-\Xop u^{k-1})\|^2_2, \\
    \rho_1^k / 2 & \text{if } \|\rho_1^k(\Xop u ^{k}-\Xop u ^{k-1})\|^2 > 10 \|\psi_1^k-\Xop u^k\|^2,  \\
    \rho_1^k     & \text{otherwise},
  \end{cases}
\end{aligned}
\end{equation}
where $\rho_1^0=0.5$.
Updates for the parameter $\rho_2$ are similar to \eqref{Eq:penalty_update}; substituting $\rho_2$ for $\rho_1$, $\nabla$ for $\Xop$, and $\psi_2$ for $\psi_1$.
Manual adjustment of parameters $\rho_1,\rho_2$ for different datasets is not needed with this approach.}
We also optimize evaluation of the \mod{X-ray} transform $\Xop$ since in the proposed framework this operator requires a relatively large amount of computational resources. Working with the data measured with the interlaced angular scanning protocol, the number of angles typically exceeds the object sizes. Therefore, instead of direct summation over the lines in computing the \mod{X-ray} transform we employ the Fourier-based method~\cite{Beylkin:95}. \mod{With this method, the forward and adjoint projection operators are evaluated faster by leveraging the speed advantage of computing 1D FFTs on regular grids with 2D FFTs on irregular grids.} For $N_\theta$ projections angles and $N$ pixels object size in one dimension, the computational complexity $\mathcal{O}(N^3\log N)$ of the Fourier-based method is sufficiently lower than the complexity $\mathcal{O}(N^3N_\theta)$ of the direct summation over lines. Alternatively, the complexity level can be decreased by employing hierarchical decomposition~\cite{George:07} or the log-polar-based method~\cite{Andersson:16}.

Third, in combination with gradual increasing optical flow resolution with iterations, we also propose a gradual increase of the recovered object resolution. Specifically, we perform a certain number of iterations for binned data and extrapolate the recovered object with optical flow and additional variables to a twice dense grid. The result is then used as an initial guess for the next set of iterations where the data are less binned. Final object and optical flow are both found on dense grids without binning. We employed this scheme for reconstructing the mouse brain, medical mask, and biological samples used in the paper. \mod{It was experimentally observed that the following strategy for automatic tuning of resolution level parameters in the proposed GPU implementation is efficient in practice. The number of binning levels is computed as $\max(1,\lceil\log_2(N/128)\rceil)$, where $N$ is the minimimum between the projection width and height. The initial optical flow window size on each binning level should be equal to the minimum between the projection width and height after binning. Then on each iteration window sizes are decreased by a small value (e.g., 1 or 2) depending on the total number of requested iterations. The number of iterations per each level is calculated in a way that the final optical flow window size on a level is twice bigger than the initial window size on the next level.}

The last important aspect involves the use of high-performance computing resources. Both deformation estimation and tomography problems are data intensive and can be parallelized with different data-partitioning methods.
Specifically, data for the deformation estimation procedure can be partitioned based on the rotation angles, and then each angle/partition can independently be processed by using the deformation operator. The tomography problem shows similar properties, where parallelization can be accomplished by slicing the object across the beam direction.

In this work, we use GPUs with Nvidia CUDA technology for accelerating computations. CUFFT and NPP libraries from CUDA Toolkit 10.1 were used for computing Fourier transforms and applying optical flow deformation with high accuracy. \mod{For fast evaluation of the forward and adjoint projection operators by using a Fourier-based method~\cite{Beylkin:95} we used CUFFT library for performing a batch of 1D and 2D FFTs on regular grids, and implemented CUDA C++ raw kernels for convolution-like procedures to interpolate data between regular and irregular grids. A multi-GPU implementation of this method can be found at \url{https://github.com/nikitinvv/usfftrecon}.} For optical flow estimation by using Farnenack's algorithm we utilized GPU-accelerated OpenCV library. Linear algebra operations in the ADMM schemes were implemented in Python using GPU-accelerated CuPy library.

In Table~
\ref{tab:perf} we provide computational times of 1 ADMM iteration for processing data sets of different sizes. To satisfy the Nyquist sampling criterion for tomography, we chose the number of angles as 3/2 of the object size in one dimension. There are several observations from this table. First, it confirms computational complexity $\mathcal{O}(N_\theta N^2)$ and $\mathcal{O}(N^3\log N)$ for the deformation estimation and tomography sub-problems, respectively, thus allowing users to estimate the whole processing time for other data sizes. 
Second, the tomographic part of the solver is more time-consuming than the deformation part and should be used as a key factor for determining processing times. However, for the interlaced scanning protocol, where the number of angles is typically much higher than the object size in one dimension, the deformation part may become more computationally demanding and crucial in estimating processing times.
Finally, we observe that performance gain with increasing the number of GPUs is not linear, specifically, it does not exceed a factor of 2 when using 4 GPUs compared to 1 GPU. This is caused by a huge amount of semi-sequential CPU-GPU data transfers. To overcome this problem, one could overlap memory transfers and computations with making use of CUDA streams. 
We plan to address this problem, so as perform other code optimization, in our further work.

A multi-GPU implementation of the proposed ADMM scheme for solving the tomography problem with projection data alignment and regularization is available at \url{https://github.com/nikitinvv/tomoalign}.

\section{Experimental data reconstruction}
In the following, we show reconstructions of projection data from two nano-tomography experiments conducted at the Advanced Photon Source in Argonne National Laboratory with the Transmission X-ray Microscope (TXM) of sector 32-ID~\cite{de2016nanoscale}.
\mod{Besides reconstructions by the proposed method we also check results obtained by a method operating with 3D optical flow and used for processing data from nano-CT synchrotron tomography experiments~\cite{odstrcil2019ab}. Its Matlab GPU-based implementation is publicly available at~\cite{odstrcil_michal_2019_2578796}. The method operates with a set of parameters that need to be tuned for getting best quality reconstructions. The parameters list includes initial and final DVF smoothness levels in spatial and temporal variables, regularizer for deformation evolution, number of pixels to describe one pixel of DVF, regularizer for preventing amplification of noise in DVF inside of empty regions, and mask constants for tomography reconstruction. There is no general guidance for choosing some of these parameters, therefore we run the solver for many different parameter combinations and select best-quality reconstructions for demonstration.
For additional comparison, our method is also tested on synthetic and experimental data available through the same source.
}

\subsection{Mouse brain data set}
\begin{figure*}[htb]
    \centering
    \includegraphics[width=1\textwidth]{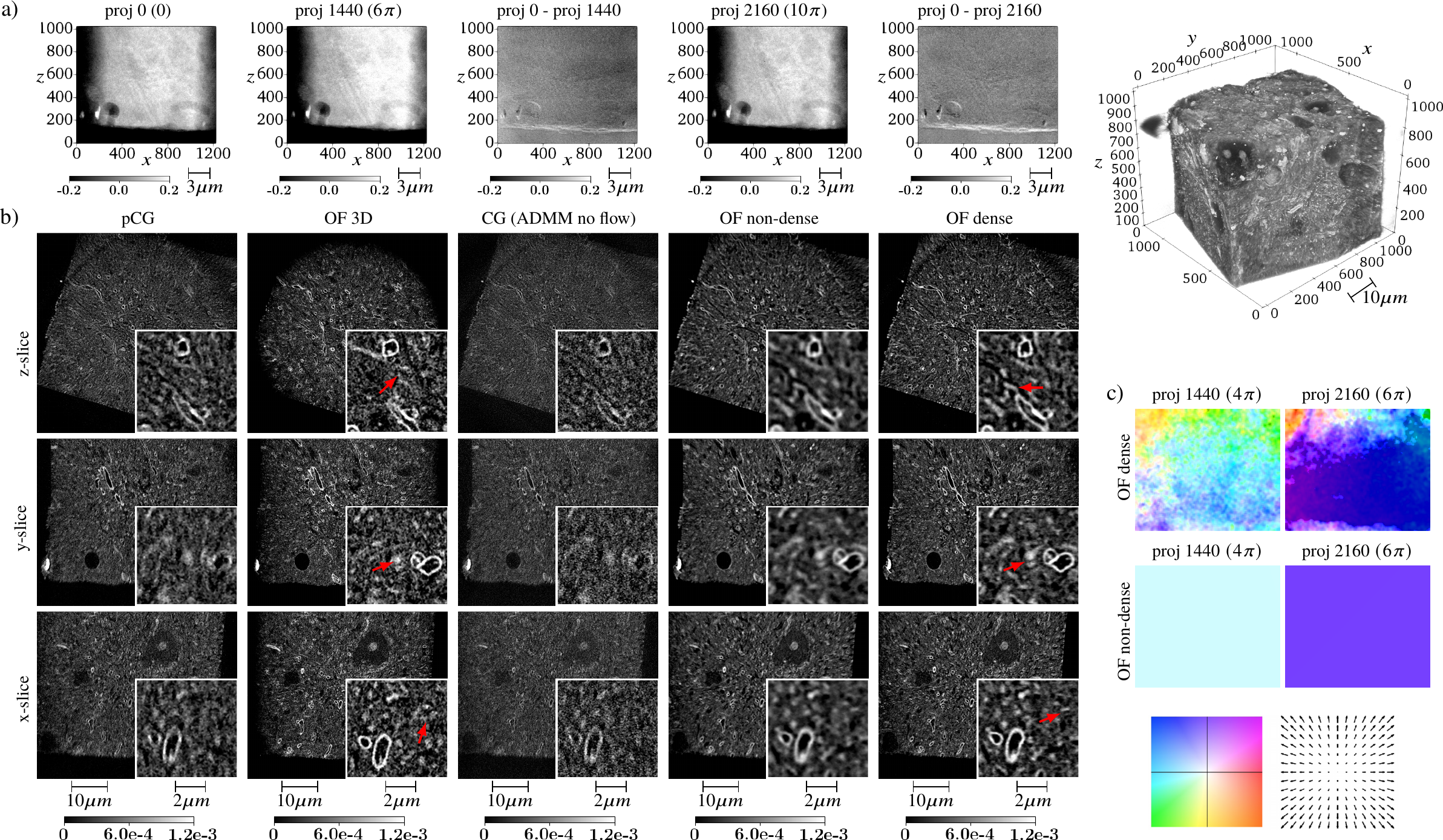}
    \caption{Reconstruction of the mouse brain absorption contrast data from the 32-ID nano-tomography beamline: a) projection data for interlaced angles and misalignment, b) reconstruction by the conventional Conjugate Gradient method \mod{(CG), its pre-aligned version (pCG)}, \mod{method with 3D deformation estimation (OF 3D)}, and by the proposed ADMM scheme with dense optical flow (OF dense) and non-dense optical flow (OF non-dense) equivalent to standard $(x,y)$-shifts of projections; \mod{Note that the OF 3D method recovers the initial object state, while other methods recover an object state corresponding to better projection consistency. Red arrows indicate examples of small structures that are noisy and corrupted in reconstruction by OF 3D, but well resolved with the OF dense method.} c) Examples of estimated dense and non-dense optical flows.}
    \label{Fig:brain}
\end{figure*}

\begin{figure*}[htb!]
    \centering
    \includegraphics[width=1\textwidth]{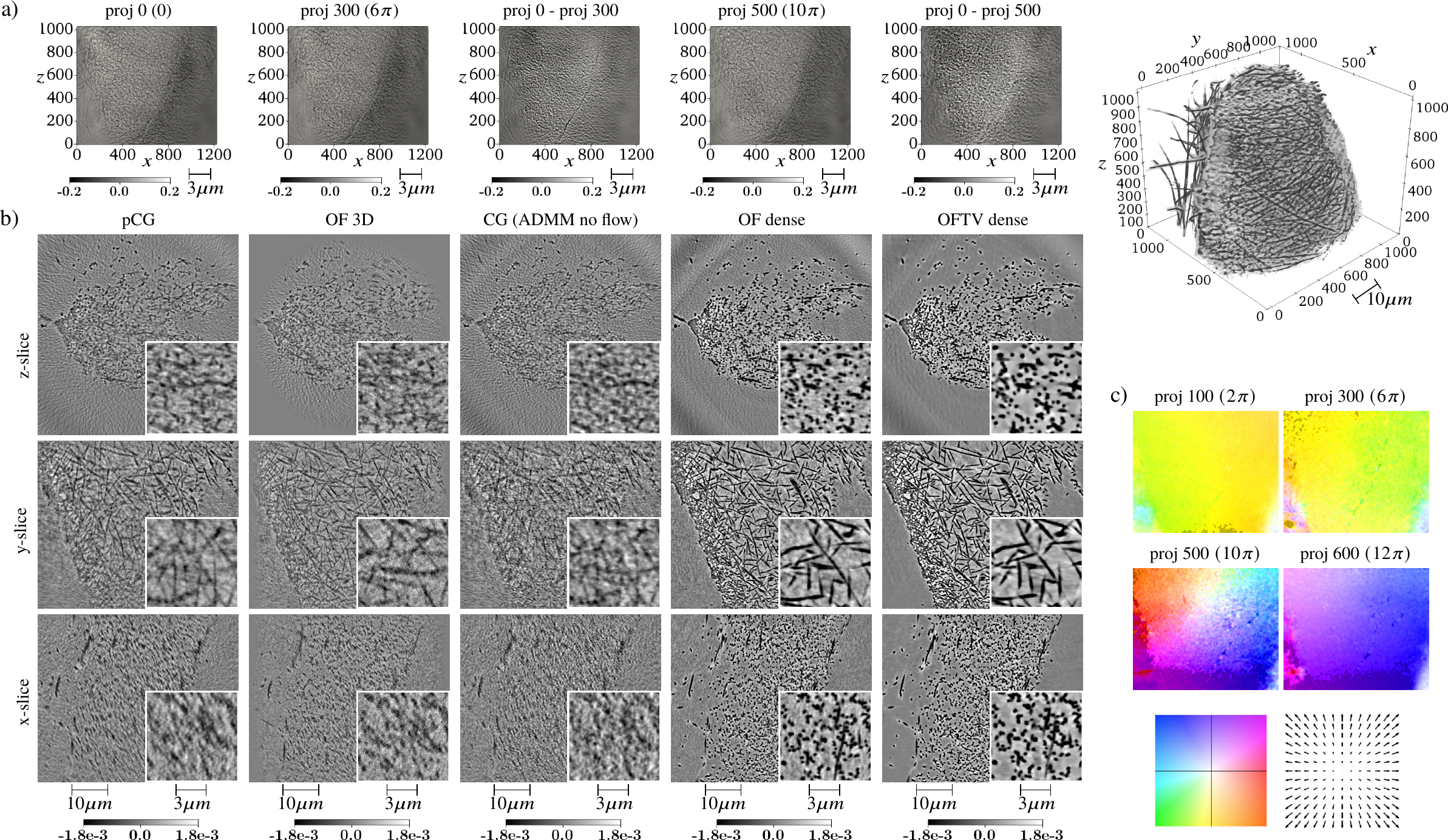}
    \caption{Reconstruction of the medical mask phase contrast data from the 32-ID nano-tomography beamline: a) projection data for interlaced angles and misalignment, b) reconstruction by the conventional Conjugate Gradient method \mod{(CG), its pre-aligned version (pCG)}, \mod{method with 3D deformation estimation (OF 3D)}, by the proposed ADMM scheme with dense optical flow (OF dense) and its TV-regularized version (OFTV dense). \mod{Note that the OF 3D method recovers the initial object state, while other methods recover an object state corresponding to better projection consistency.} c) Examples of estimated dense optical flows for interlaced projections.}
    \label{Fig:mask}
\end{figure*}
\begin{figure*}[htb!]
    \centering
    \includegraphics[width=0.9\textwidth]{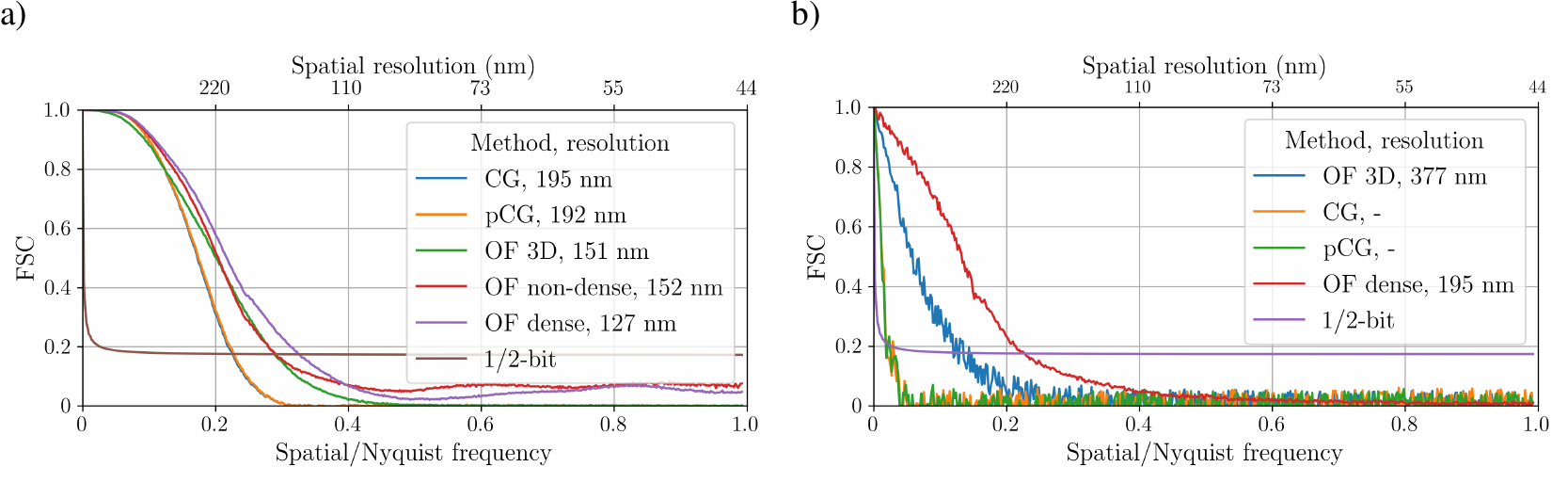}
    \caption{Fourier shell correlation (FSC) computed by using reconstructions from two independent sets of projections by the conventional Conjugate Gradient method \mod{(CG), its pre-aligned version (pCG)}, \mod{method with 3D deformation estimation (OF 3D)}, and by the proposed ADMM scheme with dense optical flow (OF dense) and non-dense optical flow (OF non-dense) for: a) mouse brain sample, b) medical mask sample.  }
    \label{fig:fsc}
\end{figure*}
The first experimental data set consists of fragments of a mouse somatosensory (S1) cortex containing myelinated axons. All animal procedures were performed in accordance with the University of Chicago Institutional Animal Use and Care Committee. The sample has been stained with heavy metals (osmium, lead)~\cite{hua2015large} to increase X-ray absorption contrast in the projections. After staining, the sample was embedded in petropoxy resin to make it more X-ray resistant.
In order to mitigate sample deformation caused by radiation damage, we employed the interlaced scanning protocol with low exposure times per projection. 
Projections were acquired in the step-scan mode at 8\,keV  with 0.2\,s exposure time per projection. In total we measured 2880 projections over an angular range of 720 degrees. The detector assembly comprises a LuAG:Ce scintillator, a 5$\times$ long working distance Mitutoyo objective lens and a FLIR CCD of the model Grasshopper3 5.0 MP Mono USB3 Vision (Sony Pregius IMX250). The CCD chip is made of $2448 \times 2048$ pixels\textsuperscript{2} with 3.45\,$\mu$m pixel size. The X-ray objective lens of the TXM is a 150\,$\mu$m large diameter 50\,nm outermost zone width Fresnel zone plate. With a CCD located 3.4\,m downstream the sample, it offers a magnification of 56 and an effective pixel size of 22.3\,nm. \mod{Since the resolving power of the Fresnel zone plate is equal to its outermost zone width, we applied $2\times 2$ data binning to decrease the amount of computations and required memory, and present reconstructions with a voxel size of 44.6\,nm}.

In Fig.~\ref{Fig:brain} we provide reconstruction results for the mouse brain data set \mod{by the conventional CG solver \textbf{(CG)}}, CG solver with pre-aligned projections \textbf{(pCG)}, \mod{the Matlab GPU-based solver with 3D deformation estimation~\cite{odstrcil_michal_2019_2578796} (\textbf{OF 3D})}, and by the proposed ADMM scheme with dense optical flow \textbf{(OF dense)}, and non-dense optical flow \textbf{(OF non-dense)} equivalent to standard $(x,y)$-shifts of projections. \mod{It should be noted that the conventional CG-solver is equivalent to the proposed ADMM solver without optical flow estimation.} In Fig.~\ref{Fig:brain}(a) we show examples of interlaced projections \mod{and} misalignment between them. Fig.~\ref{Fig:brain}(b) presents orthogonal slices through reconstructions by different methods. \mod{One can observe that reconstruction quality for the CG solver is slightly improved when projections are prealigned, whereas the proposed method demonstrates significant quality improvement and favorable for segmentation results.} At the same time, reconstruction with the dense optical flow gives sharper results than the one with non-dense optical flow, resulting in separation of even smaller features. We also note that the recovered non-dense flow is, in fact, approximately equal to the average value of the dense flow, see Fig.~\ref{Fig:brain}(c). \mod{High quality results are also demonstrated by the method based on 3D optical flow estimation. Myelinated axons (white circles) are sharp and can be easily segmented, although small structures such as the ones marked with red arrows in Fig.~\ref{Fig:brain}(b) are corrupted.} Resolution levels were estimated by computing Fourier shell correlation (FSC)~\cite{van2005fourier} between reconstructions obtained from two independent sets of 1440 projections. \mod{According to the 1/2-bit criterion, the CG and pCG methods demonstrate 195 nm and 192 nm resolution levels, respectively, which is not acceptable for further segmentation procedures. In contrast, high-resolution results of the optical flow based methods can be used for further data analysis. Demonstrated FSC resolution levels for 3D OF, OF non-dense, and OF dense methods are 151, 152, and 127 nm, respectively.} Corresponding FSC plots are placed in Fig.~\ref{fig:fsc}(a).

Projections in the OF method with dense and non-dense optical flows were aligned jointly with tomographic reconstruction. Resolution for the dense optical flow so as for the recovered object was gradually increased during iterations, \mod{following the strategy described in Section~\ref{Sec:optim}}.   Specifically, we performed 96 ADMM iterations with \num{8 x 8} data binning, followed by 48 iterations with \num{4 x 4} binning, and 24 iterations with \num{2 x 2} binning, where ADMM variables on each binning level were extrapolated to the next level sizes. The optical flow was computed starting with \num{256 x 256}, \num{128 x 128} and \num{64 x 64} window sizes in Farneback's algorithm for three binning levels, respectively, along with decreasing these window sizes by 2 on each ADMM iteration. So the final iteration for the object size $1224\times 1224\times 1024$ was computed with aligning projections by using \num{12x12} ($64-2\cdot 24 = 12$) window size in Farneback's algorithm. With the proposed scheme, the total time for OF reconstruction on 4 Tesla P100 GPUs was approximately 3 hours, whereas pCG reconstruction with 128 iterations with \num{2 x 2} binning took 2 hours. \mod{Reconstruction by the method based on 3D optical flow estimation was executed in two steps. First, the flow was estimated with 30 registration iterations  for $4\times4$ binned data (binning is required in order to fit data into GPU memory). Each iteration contained deformation estimation and 5 iterations of the tomography solver by the Simultaneous Algebraic Reconstruction Technique (SART)~\cite{andersen1984simultaneous}. In the second step, the recovered flow is extrapolated and used for reconstructing data by 30 SART iterations with binning $2\times 2$ (size of the recovered volume is $1224\times 1224\times 1024$).  The best quality reconstruction was obtained for the case with 4 subtomogram blocks representing different time frames. The whole reconstruction procedure took 4 hours.}

\subsection{Filter material for N95 grade medical masks data set}\label{Sec:filter}
The second experimental data set is related to the manufacturing of reusable N95 grade filter medium for medical masks and respirators.
The current N95 filter materials contain coarse micrometer size fibers and rely on electrostatic charge to remove harmful contaminants from the air. Any washing of this material would destroy the electrostatic charge and thereby compromise the filters protective nature~\cite{dj2016effect}. However, during an epidermic or pandemic (like the COVID-19 pandemic), there can be a significant shortfall of medical masks. Thus, the design and manufacture of reusable filter media and masks would mark a significant advancement in protecting first-line medical personnel. Nanofiber-based media fabricated using electrospinning technology can be a good candidate in this regard due to their finer nanometer-size fibers~\cite{xue2019electro}. 
Nano-tomography is used to characterize in 3D this fine electrospun media in order to understand the influence of medium microstructure on the filtration efficiency and to eventually optimize medium microstructural configurations to achieve a high efficiency of $>$95\%.

The studied material is exclusively composed of polymer fibers showing low attenuation coefficients with hard X-rays. Therefore, Zernike phase contrast ~\cite{zernike1935phase,rudolph1990amplitude,su2020x} has been employed to obtain sufficient imaging contrast on this low absorbing material. It necessitates the use of an optical mechanism consisting in  a phase ring placed in the back focal plane of the Fresnel zone plate. It translates minute variations in phase into corresponding changes in amplitude, which can be visualized as differences in image contrast. With this technique, low absorption contrast samples can be observed and recorded in high phase contrast with sharp clarity of minute specimen detail.

Projections were acquired in the interlaced step-scan mode at 8\,keV  with 1\,s exposure time per projection. In total we measured 700 projections of the size $2448 \times 2048$ pixels\textsuperscript{2}  over an angular range of 2520 degrees. \mod{\num{2 x 2} binning has been applied to projections, leading to a voxel size of 44.6\,nm, value slightly inferior the resolving power of the Fresnel zone plate with 50 nm outermost width.} 
Unlike with absorption contrast, the fiber material appears darker than the surrounding air because of the positive phase contrast effect where the phase ring selectively advances the phase of the non-diffracted beam by  $\pi/2$.

In Fig.~\ref{Fig:mask} we provide reconstruction results for the medical mask data set by the conventional CG solver with pre-alignment of projections \textbf{(pCG)}, and by the proposed ADMM scheme with dense optical flow \textbf{(OF dense)}.
Since the total number of projection angles is low, in order to get noise-free results we also equipped OF reconstructions with TV regularization, having the \textbf{OFTV dense} results. Fig.~\ref{Fig:mask}(a) shows examples of interlaced projections, so as significant misalignment between them.
Fig.~\ref{Fig:mask}(b) presents orthogonal slices through reconstructions by different methods. We observe that the conventional method in this case demonstrates completely distorted and useless results. However, the proposed OF method allows to recover an object state acceptable for further data analysis.  Additional TV regularization procedure during reconstruction helps in suppressing high-frequency noise and potentially simplifies further segmentation procedures. The method uses the TV term $\alpha\|\nabla u\|_1$ (see \eqref{Eq:minF}) with $\alpha=3\mathrm{e}{-6}$ adjusted \mod{with using the L-curve criterion~\cite{Agarwal:03}}. Examples of recovered optical flow for different projections in Fig.~\ref{Fig:mask}(c) confirm non-rigid misalignment of acquired projections. \mod{As opposed to the brain data, we were not able to obtain high quality results for the method based on 3D optical flow estimation for any algorithm parameter combinations. This is caused by significant changes between interlaced projections and angular undersampling for representing one time frame. } \begin{figure*}[htb!]
    \centering
    \includegraphics[width=0.85\textwidth]{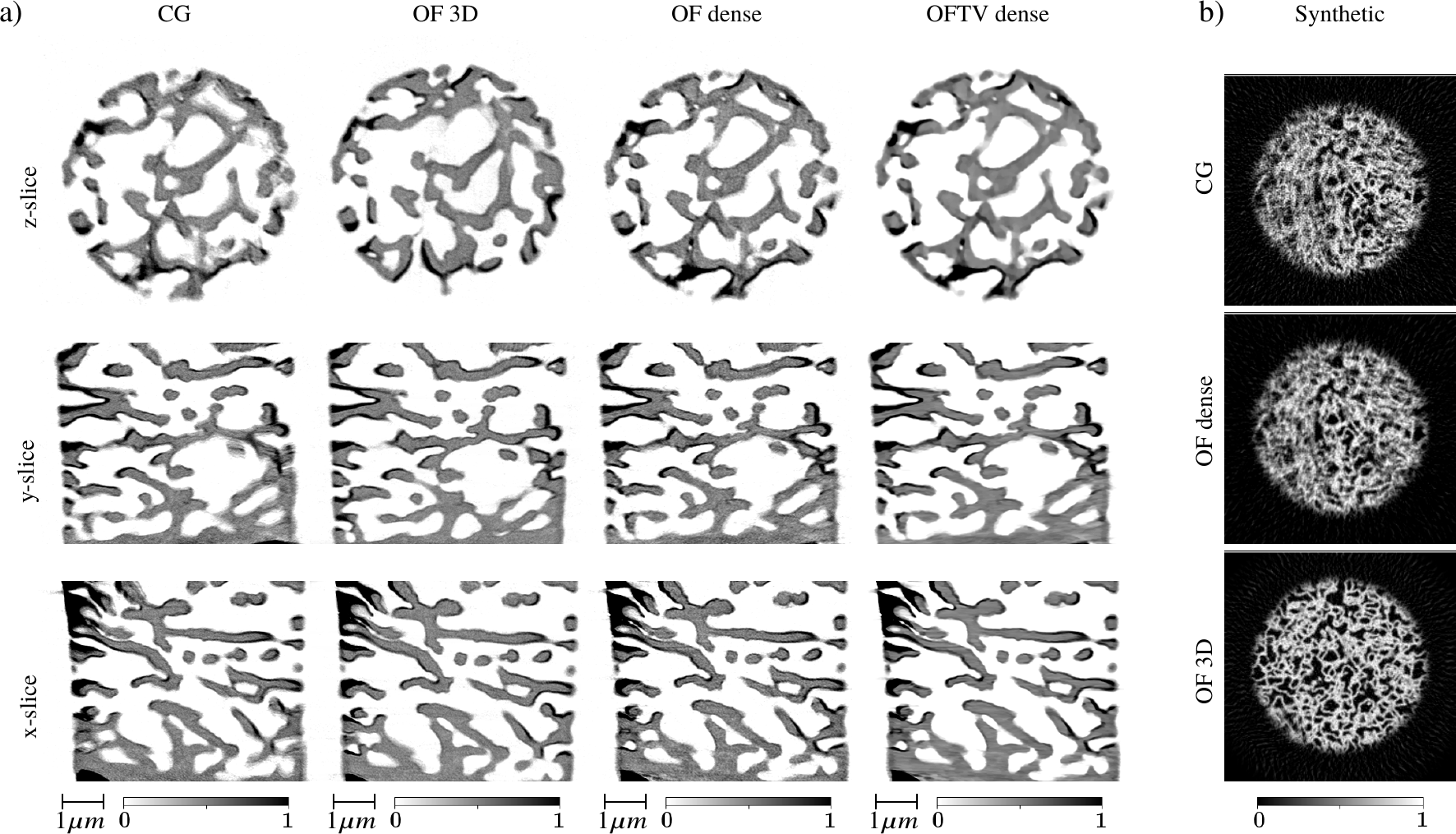}
    \caption{\mod{Reconstruction of publicly available data by the conventional Conjugate Gradient method (CG), method based on 3D optical flow estimation (OF 3D), by the proposed ADMM scheme with dense optical flow (OF dense) and its TV-regularized version (OFTV dense): a) biological sample measured with the PXCT technique, b) synthetic pillar sample of a porous material. \mod{Note that the OF 3D method recovers the initial object state, while other methods recover an object state corresponding to better projection consistency.}}}
    \label{Fig:psi}
\end{figure*}

Although the number of projections is too low, for demonstrating quantitative quality improvement we also estimate resolution levels by computing FSC between reconstructions obtained from two independent sets of 350 projections. In this case we skip the TV-regularization step in reconstruction to avoid correlation of high frequencies. According to the 1/2-bit criterion, \mod{the 3D OF and OF dense methods demonstrated 377, and 195 nm resolution levels, respectively,} while any correlation of frequencies was not observed for \mod{the CG and} pCG methods even for $>$500 nm resolution levels, see Fig.~\ref{fig:fsc}(b).

The procedure of gradual increasing resolution during iterations is the same as the one for the brain data set. The total time for OF reconstruction on 4 Tesla P100 GPUs was 1.8 hours, whereas reconstruction \mod{by the CG method} with 128 iterations took 1.5 hours. \mod{As for the brain data, two steps were performed for the method based on 3D optical flow. The flow was estimated with 30 registration iterations for $4\times4$ binned data. Each iteration contained deformation estimation and 5 iterations of the tomography solver by the SART method. Then the recovered flow is used for reconstructing data by 30 SART iterations with no binning. The best quality reconstruction was obtained for the case with 3 subtomogram blocks representing different time frames. The whole procedure took 1.2 hours. }

\mod{\subsection{Publicly available data}\label{Sec:public}
In this section, we present additional comparison between the proposed method and the method based on 3D optical flow on two data sets publicly available at~\cite{odstrcil_michal_2019_2578796} and used for demonstrating high-quality reconstructions in~\cite{odstrcil2019ab}. The first data set was acquired by scanning a biological sample with the ptychographic X-ray computed tomography (PXCT) technique~\cite{dierolf2010ptychographic}. Projections were recovered with a two-dimensional ptychographic solver, and used as input for the tomographic solver with deformation compensation. In total, there are 380 projections of the size $544\times320$, where angles are presented by three 180 degrees intervals. The unmodified Matlab script for 3D optical flow estimation taken from ~\cite{odstrcil_michal_2019_2578796} operates with $4\times4$ data binning for deformation estimation, therefore total reconstruction time is low and equal to 1.5 min.  
In turn, 4 min was needed to reconstruct data by the proposed method with automatic parameters tuning described in Section~\ref{Sec:optim} and dense optical flow estimation, although  high quality reconstructions can be obtained much faster with lower optical flow density. Fig.~\ref{Fig:psi}(a) presents orthogonal slices through the volumes reconstructed by the conventional CG solver (\textbf{CG}), the Matlab GPU-based solver with 3D deformation estimation \textbf{OF 3D}, the proposed ADMM solver with dense optical flow (\textbf{OF dense}), and its regularized version \textbf{OFTV dense}. Significant reconstruction quality improvement is demonstrated with both OF 3D and OFTV dense methods. Note that reconstructions look different because OF 3D recovers the initial object state, but OFTV dense recovers a state that corresponds to better projection consistency.}

\mod{
The second data set is synthetic. Projections were generated for 380 interlaced angles from a simulated pillar of a porous material with dimensions of $288\times288\times120$. Reconstruction results are presented in Fig.~\ref{Fig:psi}(b). Here we observe that the method based on 3D optical flow demonstrates perfect results, however, our proposed method fails. 
Compared to the previous data sets considered in this paper, this sample has a homogeneous porosity structure with a huge amount of pores. Deformation of a small inner structure of the object yields insignificant changes in projections. Three-dimensional optical flow is more favorable in this case because it operates with two separable intensity levels: pores (white) and air (dark). Optical flow on projections, in turn, operates with a lot of different intensities obtained with summation over rays propagating through the object. These intensities are close to each other, therefore it is almost impossible to track any motion.
}

\section{Discussions and outlook}
In this paper, we demonstrated that projection data consistency regulated by dense optical flow inside the ADMM scheme can result in reconstructing one object state with less artifacts.  The proposed scheme does not require any explicit prior knowledge about the object structure and motion. Moreover, the method has lower computational complexity and spends \mod{comparable processing time for} large-scale experimental data as the conventional CG solver.
Interlaced scanning with a low number of angles per rotation is favorable for the proposed method, however, the number of angles in real experiments should be adjusted according to overhead for rotating the stage, which affects total acquisition times and sample stability.

\mod{The proposed scheme particularly allows for compensating deformations occurred while scanning radiation sensitive samples. It should be noted that the scheme as other reconstruction schemes operating with optical flow works only when radiation damage induce only nondestructive deformations, and cannot replace the use of cryogenic systems preventing destruction of the sample. In such cases the scheme can be efficiently used in combination with these cryogenic systems to improve resolution levels in reconstructions.}

\mod{We provided a detailed comparison of the proposed ADMM scheme to a method based on 3D optical flow estimation. First difference is that our scheme improves consistency of projections to recover one pristine image of the sample with less motion artifacts, while 3D optical flow recovers dynamics of the sample at different time steps. 
Second, there are cases where both approaches may fail. For instance, 3D optical flow fails whenever deformation between recovered time frames is too fast and the number of projections per rotation is low (as in the case of reconstructing a filter material sample in Section~\ref{Sec:filter}). The proposed projection-based optical flow may fail, for instance, whenever the object has a dense structure of similar intensities and deformation occurs locally (as in the synthetic sample in Section~\ref{Sec:public}). 
Third, the case with 3D optical flow applied to the object is computationally and memory demanding because optical flow needs to be estimated for 3D volumes. In all provided experimental data, reconstructions with sizes bigger than 1000 in a single dimension, we had to apply additional binning and estimate optical flow with lower resolution. Reconstruction times are comparable only if binning is applied and the total number of tomographic iterations is significantly lower for the 3D optical flow based method. With our method we were able to operate with the data sizes exceeding 2000 in a single dimension, although such examples are not presented in this work because experimental data from TXM is generally binned before reconstruction in order to have voxel sizes close to the resolving power of the Fresnel zone plate.
Finally, the proposed ADMM scheme has favorable convergence behavior. While testing 3D optical flow estimation from
\cite{odstrcil_michal_2019_2578796}, we observed that for some parameters the algorithm diverges. In turn, convergence of the proposed ADMM implementation is controlled through coordinating terms and automatically tuned weights ($\lambda_1,\lambda_2,\rho_1,\rho_2$ in equations \eqref{Eq:tomo2}-\eqref{Eq:regul2}). Such convergence control is missing in many ad hoc solutions based on the alternating approach with two sub-problems.}

The ADMM framework is extendable with new sub-problems and additional data processing procedures. 
\mod{An immediate extension of the proposed scheme could be adding a sub-problem for 3D optical flow estimation, and this way making it possible to handle all types of the sample deformation.}
\mod{Other} procedures that are common in experimental data processing workflow may contain, for instance, data deblurring~\cite{liu2009deconvolution,koho2019fourier}, denoising and quality enhancement with machine learning~\cite{pelt2018improving,liu2020tomogan}. Technically, implementation of new sub-problems is straightforward by adding new auxiliary variables as in \eqref{Eq:minF}. It is also possible to prepare a general Plug-and-Play ADMM framework (similar to the one proposed in~\cite{venkatakrishnan2013plug}) for denoising, where state-of-the-art learning based methods can be applied into the scheme only by changing the implementation of corresponding models.

ADMM provides a generic framework by defining the sub-problems, but it does not provide a recipe for the solution of each sub-problem. For example, instead of using a CG solver, the tomography sub-problem can be solved by using second-order optimization methods with favorable convergence behavior. However, as discussed above, it is known that the convergence rate of the ADMM drops significantly when the sub-problems are solved exactly or when the convergence is achieved. Therefore, first-order or methods like CG are often employed in ADMM type solvers, because they provide approximate solutions quickly and effectively (e.g. few inner iterations instead of 100s of iterations until convergence). In addition, in practice second-order methods require heavy computation and memory usage, which restricts their use when processing large data sets. This was one of the reasons we often use gradient-descent or CG for inner ADMM iterations.

We are also working on developing a joint ptycho-tomography solver for nano-imaging of dose sensitive samples~\cite{gursoy2017direct, Aslan:19, nikitin2019photon}, and a similar ADMM scheme (with complex calculus instead of real) can easily be re-purposed for another application. As a follow-up step, we plan to merge formulations and construct a generic framework for ptycho-tomography that will also include alignment and deformation estimation procedures. We hope that the optical flow based nonrigid alignment may be useful in addressing the position correction problem~\cite{zhang2013translation}, which is known as a computationally demanding procedure. We also plan to develop a multi-GPU implementation on several computing nodes to further extend the computational efficiency of the framework.

\section*{Acknowledgments}

This research used resources of the Advanced Photon Source, a U.S. Department of Energy (DOE) Office of Science User Facility operated for the DOE Office of Science by Argonne National Laboratory under Contract No. DE-AC02-06CH11357.
The research was also partly supported by the Marie Sklodowska-Curie Innovative Training Network MUMMERING  (Multiscale, Multimodal, Multidimensional imaging for EngineeRING), funded  through the EU research programme Horizon 2020 and by the European Research Council through Consolidator Grant No. 681881.

\bibliographystyle{IEEEtran}
\bibliography{refs}

\end{document}